\documentclass[aps, twocolumn, letterpaper, superscriptaddress, prl, showpacs]{revtex4}

\usepackage{ifthen}
\usepackage{graphicx}
\usepackage{amsmath}
\usepackage{revsymb}

\def\A{{\bf A}}
\def\B{{\bf B}}
\def\Be{{\bf B}^{\text{ext}}}
\def\Bi{{\bf B}^{\text{ind}}}
\def\bi{B^{\text{ind}}}
\def\be{B^{\text{ext}}}

\def\F{{\bf F}}
\def\G{{\bf G}}

\def\M{{\bf M}}
\def\m{{\bf m}}

\def\p{{\bf p}}
\def\rvec{{\bf r}}
\def\R{{\bf R}}

\def\emac{\cal E}
\def\Emac{\boldsymbol{\cal E}}

\begin{document}

\title{A converse approach to the calculation of NMR shielding tensors}
\author{T. Thonhauser}
\affiliation{Department of Physics, Wake Forest University, Winston-Salem,
North Carolina 27109, USA.}
\affiliation{Department of Materials Science and Engineering,
MIT, Cambridge, Massachusetts 02139, USA.}
\author{D. Ceresoli}
\affiliation{Department of Materials Science and Engineering,
MIT, Cambridge, Massachusetts 02139, USA.}
\author{Arash A. Mostofi}
\altaffiliation{Current address: Department of Materials, Imperial College
London, SW7 2AZ, United Kingdom.}
\affiliation{Department of Materials Science and Engineering,
MIT, Cambridge, Massachusetts 02139, USA.}
\author{Nicola Marzari}
\affiliation{Department of Materials Science and Engineering,
MIT, Cambridge, Massachusetts 02139, USA.}
\author{R. Resta}
\affiliation{Dipartimento di Fisica Teorica Universit\`a di Trieste and
INFM-DEMOCRITOS, Trieste, Italy.}
\author{David Vanderbilt}
\affiliation{Department of Physics and Astronomy, Rutgers University,
Piscataway, New Jersey 08854, USA.}
\date{\today}

\begin{abstract}  
We introduce an alternative approach to the first-principles calculation
of NMR shielding tensors. These are obtained from the derivative of the
orbital  magnetization with respect to the application of a microscopic,
localized magnetic dipole. The approach is simple, general, and can be
applied to either isolated or periodic systems. Calculated results for
simple hydrocarbons, crystalline diamond, and liquid water show very
good agreement with established methods and experimental results.
\end{abstract}


\maketitle

Nuclear magnetic resonance (NMR) measures the transition frequencies for
the reorientation of nuclear magnetic moments in an applied magnetic
field. Since the local magnetic field differs from the external one as a
result of electronic screening, NMR spectroscopy~\cite{NMR_encyclopedia}
has been recognized since 1938~\cite{Rabi} to be a powerful experimental
probe of local chemical environments, including structural and
functional information on molecules, liquids, and increasingly, on
solid-state systems.

First-principles calculations of NMR spectra were first developed in the
quantum chemistry community~\cite{Kutzelnigg_90} and applied to
molecules and clusters, but applications to extended crystalline systems
were hindered by the difficulty of including macroscopic
magnetic fields, which require a non-periodic vector potential
that is not compatible with Bloch symmetry. In 1996 Mauri
\emph{et al.}\ developed a linear-response approach for calculating NMR
shieldings in periodic crystals based on the long-wavelength limit of a
periodic modulation of the applied magnetic field~\cite{Mauri_96}.
In principle, this long-wavelength approach could be implemented using
standard ground-state calculations, but only at the cost of introducing
prohibitively large supercells. In 2001 Sebastiani and Parrinello used
a localized Wannier representation~\cite{Marzari_97} to derive an
alternative linear-response approach based on the application of an
infinitesimal uniform magnetic field~\cite{Sebastiani_01}. More recently,
attention has focused on the development of these approaches in the
context of pseudopotentials~\cite{Pickard_Mauri_01,Yates_07},
relying on ideas developed within the projector-augmented wave method
(PAW)~\cite{Blochl_94}, leading to a growing use of these methods in
combination with modern plane-wave pseudopotential codes
\cite{paratec, castep}. Despite these advances, existing methods for
computing NMR shifts in crystalline systems remain complex, in that
they require a linear-response implementation with significant extra
coding.

In this Letter, we reformulate the problem of computing NMR shielding
tensors so that the need for a linear-response framework is
circumvented.  For clarity, the previous formulations shall be referred
to as \emph{direct} approaches, in that a magnetic field is applied and
the local field at the nucleus is computed. Our alternative,
\emph{converse} approach obtains the NMR shifts instead from the
macroscopic magnetization induced by magnetic point dipoles placed at
the nuclear sites of interest. This approach is made possible in
periodic systems by the recent developments that have led to the
Berry-phase modern theory of magnetization
\cite{Resta_05,Thonhauser_06,Ceresoli_06,Niu}. We demonstrate the method
by a first application to molecules, crystals, and liquids, finding
excellent agreement both with experiment and with calculations using the
direct approach.  Our new method is simple and general, and provides a
straightforward alternative avenue to the computation of NMR shifts,
which is suited for large-scale simulations and situations where a
linear-response formulation is cumbersome or unfeasible.

Let us start by considering a sample to which a constant external
magnetic field $\Be$ is applied. The field induces a current that, in
turn, induces a magnetic field $\Bi(\rvec)$ such that the total magnetic
field is $\B(\rvec) = \Be  + \Bi(\rvec)$.  In NMR experiments the
applied fields are small compared to the typical electronic scales;  the
absolute chemical shielding tensor $\tensor{\sigma}$ is then defined via
the linear relationship
\begin{equation}\label{equ:sigma}
\Bi_s=-\tensor{\sigma}_s\cdot\;\Be\;,
\quad \sigma_{s,\alpha\beta} =
-\frac{\partial\bi_{s,\alpha}}{\partial \be_\beta}\;.
\end{equation}
The index $s$ indicates that the corresponding quantity is to be taken
at position $\rvec_s$, i.e., the site of nucleus $s$. NMR experiments
usually report the isotropic shielding
$\sigma_s=\frac{1}{3}{\text{Tr}}[\,\tensor{\sigma}_s\,]$ via a chemical
shift that is defined by convention as $\delta_s =
\sigma_{\text{ref}}-\sigma_s$, where $\sigma_{\text{ref}}$ is the 
isotropic shielding of a reference compound.

As mentioned above, direct approaches
\cite{Mauri_96,Sebastiani_01,Pickard_Mauri_01,Yates_07} calculate the
chemical shielding from the current response of the system to an
external magnetic field, applied using perturbation theory and taking
the long-wavelength limit.  The approach we propose is fundamentally
different: instead of determining the current response to a magnetic
field, we derive chemical shifts from the orbital magnetization induced
by a magnetic dipole. This can be shown using a thermodynamic
relationship between mixed partial derivatives \cite{Ditchfield_72}, as
follows: Using $B_{s,\alpha}=\be_\alpha + \bi_{s,\alpha}$,
Eq.~(\ref{equ:sigma}) becomes $\delta_{\alpha\beta} -
\sigma_{s,\alpha\beta} = \partial B_{s,\alpha}/\partial\be_\beta$. For
the moment, we assume that $\Be$ can be replaced by the total
macroscopic $B$-field in the denominator of this equation, thus
neglecting the {\it macroscopic} induced field (this restriction,
appropriate for normal components in a slab geometry, will be relaxed
shortly). The numerator may be written as $B_{s,\alpha} = -\partial
E/\partial m_{s,\alpha}$, where $E$ can be interpreted either as the
energy of a virtual magnetic dipole $\m_s$ at one nuclear center
$\rvec_s$ in the field $\B$ for a finite system, or as the energy per
cell of a periodic lattice of such dipoles; we adopt the latter view. 
Then, writing the macroscopic magnetization as
$M_\beta=-\Omega^{-1}\,\partial E/\partial B_\beta$ (where $\Omega$ is
the cell volume), we obtain
\begin{equation}\label{equ:converse}
\delta_{\alpha\beta} - \sigma_{s,\alpha\beta} =
  - \frac{\partial}{\partial B_\beta}
  \frac{\partial E}{\partial m_{s,\alpha}} =
  - \frac{\partial}{\partial m_{s,\alpha}}
  \frac{\partial E}{\partial B_\beta} =
  \Omega\frac{\partial M_\beta}{\partial m_{s,\alpha}}\;.
\end{equation}
Thus, $\tensor{\sigma}_s$ accounts for the shielding contribution to the
macroscopic magnetization induced by a magnetic point dipole $\m_s$
sitting at nucleus $\rvec_s$ and all of its periodic replicas. In other
words, instead of applying a constant (or long-wavelength) field $\Be$
to an infinite periodic system and calculating the induced field at all
equivalent nuclei $s$, we apply an infinite array of magnetic dipoles to
all equivalent sites $s$ and calculate the change in orbital
magnetization \cite{Resta_05,Thonhauser_06,Ceresoli_06,Niu}. Since the
perturbation is now periodic, it can easily be computed using finite
differences of ground-state calculations. This is our principal result. 
Note that $\M=\m_s/\Omega + \M^{\text{ind}}$, where the first term is
present merely because we have included magnetic dipoles by hand. It
follows that the shielding is related to the true induced magnetization
via $\sigma_{s,\alpha\beta}= -\Omega\,\partial
M^{\text{ind}}_\beta/\partial m_{s,\alpha}$.

It is useful to pause here and consider the analogy with the
Born~\cite{Born}  effective charge tensor $Z^*_{s,\alpha\beta}$, which
may be regarded as (i) the component of the force $\F_s$ in direction
$\alpha$ on site $\rvec_s$ by a unit macroscopic electric field $\Emac$
in direction $\beta$ (at zero nuclear displacement), or, alternatively,
as (ii) the $\beta$-component of the  macroscopic electric polarization
${\bf P}$ linearly induced by a unit displacement of nucleus $s$ and its
periodic replicas in direction $\alpha$, in a vanishing macroscopic
electric field.  Since the force on  nucleus $s$ is given by
$F_{s,\alpha}=-\partial  E/\partial r_{s,\alpha}$, (i) and (ii) are
related by
\begin{equation}\label{equ:analogy}
Z^*_{s,\alpha\beta}
= -\frac{\partial}{\partial\emac_\beta}
  \frac{\partial E}{\partial r_{s,\alpha}} 
= -\frac{\partial}{\partial r_{s,\alpha}}
  \frac{\partial E}{\partial\emac_\beta}
=  \Omega\frac{\partial P_\beta}{\partial r_{s,\alpha}}\;,
\end{equation}
in close analogy with Eq.~(\ref{equ:converse}). Note that, in order to
comply with the Born definition, one must choose the
\emph{lattice-periodical} solution of Poisson's equation, corresponding
to vanishing macroscopic electric field. Other choices are possible and
lead to other kinds of effective charges, which are related to each
other via the dielectric constant~\cite{Ghosez98}. By comparing
Eq.~(\ref{equ:analogy}) to Eq.~(\ref{equ:converse}) we notice that the
genuine analogue to $\tensor{Z^*_s}$ is  $1 - \tensor{\sigma}_s$ (and
{\it not} $\tensor{\sigma}_s$), as indeed the names ``effective'' vs.
``shielding'' imply. As a side note, it should be pointed out that this
analogy between the electric and magnetic cases would be formally more
direct if the nucleus carried a magnetic monopole charge.

As in the electrical case~\cite{Ghosez98}, the choice of magnetic
boundary conditions implies a choice for the {\it shape} of the
macroscopic finite sample. Following Ref.~\cite{Landau1}, shape effects
can be embedded in the depolarization coefficients ${n}_\alpha$ (with
$\sum_\alpha {n}_\alpha=1$), whose special cases are the sphere
($n_x$=$n_y$=$n_z$=1/3), the cylinder along $z$ ($n_x$=$n_y$=1/2,
$n_z$=0), and the slab normal to $z$ ($n_x$=$n_y$=0, $n_z$=1). The main
relationship for the macroscopic fields in Gaussian units may be written
as $B_\alpha = \be_\alpha + 4 \pi (1 - n_\alpha) M_\alpha$. It can be
seen that for the slab geometry the normal component of $\B$ coincides
with the one of $\Be$. Hence our computed $\sigma_{s,zz}$ are suitable
for direct comparison with measurements of the normal component
performed on a slab-shaped sample. Assuming non-magnetic media with
small, isotropic susceptibility $\chi$, it can be shown that the
shielding for a general shape is related to our  calculated one by
$\sigma_{s,\alpha\beta}^{\text{shape}}\simeq\sigma_{s,\alpha\beta} - 
\delta_{\alpha\beta}\,4\pi\chi(1-n_\beta)$. For the special case of a
spherical sample we have
$\sigma_{s,\alpha\beta}^{\text{sphere}}\simeq\sigma_{s,\alpha \beta} -
(8\pi/3)\,\chi\,\delta_{\alpha\beta}$.

In order to calculate the shielding tensor of nucleus $s$ using
Eq.~(\ref{equ:converse}), it is necessary to calculate the induced
orbital magnetization due to the presence of an array of point magnetic
dipoles $\m_s$ at all equivalent sites $\rvec_s$. The vector potential
of a single dipole in Gaussian units is given by~\cite{Jackson_dipole}
\begin{equation}\label{equ:A_sr}
\A_s(\rvec) = \frac{\m_s \times (\rvec - \rvec_s)}
{ |\rvec - \rvec_s|^3} \;.
\end{equation}
For an array of magnetic dipoles $\A(\rvec)=\sum_{\R}\A_s(\rvec-\R)$,
where $\R$ is a lattice vector. Since $\A$ is periodic, the average of
its magnetic field $\nabla\times\A$ over the unit cell vanishes; thus,
the eigenstates of the Hamiltonian remain Bloch-representable. In the
Fourier representation $\A(\rvec)=\sum_{\G\neq0}\tilde{\A}(\G)\,{\rm
e}^{i\G\cdot\rvec}$ with
\begin{equation}
\tilde{\A}(\G) = -\frac{4 \pi i}{\Omega} \frac{\m_s \times
 \G}{G^2}\,{\rm e}^{-i \G \cdot \rvec_s},
\end{equation}
where the reciprocal lattice vector $\G=0$ may be excluded from the sum
with no loss of generality. Note that we have implicitly chosen the
transverse gauge $\nabla\cdot\A=0$, which is apparent from
$\G\cdot(\m_s\times\G)=0$. The periodic vector potential $\A(\rvec)$ can
now be included in the Hamiltonian with the usual substitution for the
momentum operator $\p\rightarrow \p-\frac{e}{c}\A$. As a result, the
kinetic energy operator becomes
\begin{equation}\label{equ:new_terms}
\frac{p^2}{2m_{e}} \longrightarrow
\frac{p^2}{2m_{e}} - \frac{e}{m_ec}\A\cdot\p + 
\frac{e^2}{2m_ec^2}A^2\;, 
\end{equation}
where $m_{e}$ is the electronic mass and $c$ is the speed of light. Due
to our choice of gauge, $\p$ and $\A$ commute.  We can now calculate the
shielding according to Eq.~(\ref{equ:converse}) by solving for the
ground state with the additional terms of Eq.~(\ref{equ:new_terms})
included in the Hamiltonian \cite{A2}, and then calculating the
resulting change in orbital magnetization.

The converse method can be implemented directly in any all-electron
electronic-structure code. However, many popular density-functional
theory codes use pseudopotentials to increase computational efficiency.
In order to calculate NMR shifts in the presence of pseudopotentials, a
PAW reconstruction \cite{Blochl_94} needs to be performed, as shown by
Pickard and Mauri \cite{Pickard_Mauri_01}. We have developed this
reconstruction methodology for the converse method; the rather involved
mathematical formalism will be presented elsewhere \cite{Ceresoli_09}.
We implemented our converse approach, including this reconstruction and
the calculation of orbital magnetization 
\cite{Resta_05,Thonhauser_06,Ceresoli_06,Niu}, into the {\sc PWscf}
package of the {\sc Quantum-ESPRESSO} distribution \cite{pwscf}. We use
the PBE exchange-correlation functional \cite{PBE} and 
Troullier-Martins pseudopotentials \cite{TM}, with convergence of the
NMR shifts for a kinetic-energy cutoff of 80~Ryd. The dipole
perturbation $|\m_s|$ used is 1$\mu_\mathrm{B}$, although we find
identical results for any value in the broad range $10^{-6}$ to
$10^3$$\mu_\mathrm{B}$.

As an initial test, we applied our converse approach to the calculation
of hydrogen NMR chemical shieldings for small molecules using a
supercell geometry (in such cases the full orbital-magnetization theory
is not needed, as the magnetization can be obtained just by integrating
the orbital currents over the molecular region, but the results
presented here were in fact obtained using the Berry-phase modern theory
of magnetization). For purposes of comparison, we also calculated
these shieldings with the direct method, also implemented by some of us
in {\sc PWscf}, largely eliminating discrepancies due to any
technicality. The results for these two approaches are shown in
Table~\ref{tab:results} together with the experimental values. It is
immediately obvious that the direct and converse methods give almost
identical results, validating our approach, and also very good agreement
(we suspect that the slight deviations between the two calculations are
due to the long-wavelength approximation of the direct method).

\begin{table}
\caption{\label{tab:results} Hydrogen NMR chemical shielding $\sigma$, in ppm, for
several different molecules. Structural parameters were taken from footnote 22
of Ref.~\cite{Mauri_96}.}
\begin{tabular*}{\columnwidth}{@{}l@{\extracolsep{\fill}}ccr@{}}\hline\hline
         & experiment  & direct & converse   \\\hline
H$_2$     & 26.26$\,^a$ & 26.2   & 26.2       \\
HF        & 28.51$\,^a$ & 28.4   & 28.5       \\
CH$_4$    & 30.61$\,^a$ & 30.8   & 31.0       \\
C$_2$H$_2$& 29.26$\,^b$ & 28.8   & 28.9       \\
C$_2$H$_4$& 25.43$\,^b$ & 24.7   & 24.8       \\
C$_2$H$_6$& 29.86$\,^b$ & 30.2   & 30.4       \\ \hline\hline
\end{tabular*}
\raggedright
$^a\,$Reference \cite{Raynes}.\\
$^b\,$Reference \cite{Schneider}.
\end{table}

Next, we applied our method to crystalline diamond. For our calculations
we used an 8-atom cubic cell with a lattice constant of 3.498~\AA. The
NMR shielding converged to within 0.1~ppm for a k-point mesh of
$8\times8\times8$.  For the $^{13}$C shielding we find $131.20$~ppm, in
perfect agreement with the direct method. To estimate the effect of the
spurious interactions of the localized dipole with its images in
neighboring supercells, we repeated the calculation for a 64-atom cubic
cell, finding an almost identical shielding of $131.24$~ppm. The fast
convergence with respect to supercell size is due to the fast decay
$(1/r^2)$ of the vector potential in real space.

\begin{figure}
\includegraphics[width=0.9\columnwidth]{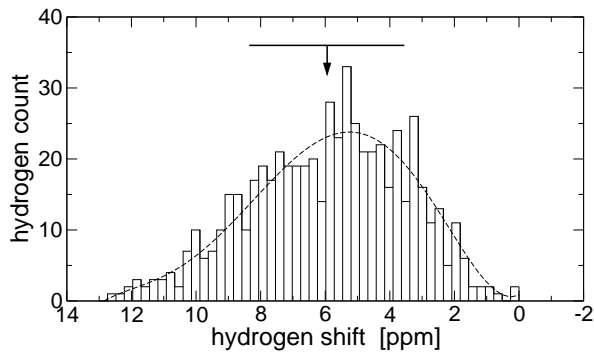}
\caption{\label{fig:water}Distribution of hydrogen NMR shifts in liquid
water relative to the gas-phase shift. The distribution was obtained
from five snapshots of the 64-molecule system (640 hydrogen atoms). The
dashed line is a polynomial fit and it serves as a guide to the eye. The
vertical arrow and horizontal line indicate the position of the average
and the range of the standard deviation, respectively.}
\end{figure}

Finally, we applied the converse approach to compute the hydrogen
chemical shifts in a supercell simulation of liquid water.  Our
supercell contained 64 water molecules, twice the size of the largest
supercell used in previous NMR calculations on liquid water using the
direct method \cite{Sebastiani_02, Pfrommer_00}. We obtained the atomic
trajectories from a molecular-dynamics simulation using TIP4P
\cite{TIP4P} potentials under standard conditions. For five snapshots
separated by 200 ps, we took the atomic positions and thermalized the
hydrogen atoms alone for 2 ps using ab-initio Car-Parrinello molecular
dynamics. This procedure is aimed at obtaining a more realistic
description of the detailed structure of the water molecules while
retaining the accuracy of the oxygen-oxygen pair correlation function. 
We then used these positions to calculate hydrogen shifts with the
converse method. We calculated the shift of liquid water with reference
to the gas-phase shift, i.e., $\delta_{\text{liquid}} =
\sigma_{\text{gas}} - \sigma_{\text{liquid}}$, thus reporting the
experimental measurable change  for the gas-liquid transition. For the
gas-phase shift the converse method gives 31.0~ppm. For the
susceptibility correction of periodic water we used the experimental
value for $\chi_{\text{water}}=-7.2\times 10^{-7}$ emu under standard
conditions \cite{CRC}. Our distribution for the hydrogen shifts, shown
in Fig.~\ref{fig:water}, can be directly compared to results obtained
using the direct method reported in Fig.~4 of
Ref.~\cite{Sebastiani_02} and in Ref.~\cite{Pfrommer_00}. 
We find an average shift of 5.94~ppm from our distribution, in excellent
agreement with 5.83~ppm and 5.15~ppm from
Refs.~\cite{Sebastiani_02} and \cite{Pfrommer_00},
respectively. Furthermore, the spread of our distribution as measured by
the standard deviation is 2.4~ppm, again in precise agreement with the
value of 2.4~ppm obtained from the direct method \cite{Pfrommer_00}.

At first sight it might appear that the converse method is
computationally more demanding than the direct method, since we need to
perform $3N$ calculations to obtain the shielding tensor for $N$ atoms.
Often, though, only a few selected shifts are needed, and even in the
worst-case scenario (such as the water calculation shown before) it
should be stressed that re-minimizing the electronic wavefunctions in
the presence of the perturbation is very fast, usually requiring a
single self-consistent iteration.  This enables the calculation of NMR
shielding tensors for systems with several hundred atoms even on small
computer clusters. However, the main advantage of the converse method is
the simplicity of its implementation, in that it works via finite
differences of ground-state calculations and does not require a
linear-response implementation.  This is likely to be a significant
advantage for future applications in conjunction with more complex forms
of exchange-correlation functionals such as DFT+U, exact exchange, 
hybrid functionals, or beyond-DFT correlated-electrons methods.

In conclusion, we have derived an alternative first-principles method
for calculating NMR chemical shielding tensors in solids.  The new
approach is considerably simpler than existing techniques, avoiding
altogether the difficulties related to the choice of a gauge origin  and
the need for a linear-response implementation. We have demonstrated the
correctness and viability of our approach by calculating chemical
shieldings in isolated and periodic systems: simple molecules,
crystalline diamond, and liquid water, finding excellent agreement with
previous theoretical and experimental results. Applications to more
complex systems, including small proteins and systems with heavy
elements, are currently in progress.

This work was supported by NSF grant DMR-0549198, ONR grant
N00014-07-1-1095, the DOE/SciDAC project on Quantum Simulation of
Materials and Nanostructures, and MITEI/ENI. All computations were
performed on the WFU DEAC Cluster with support from the WFU Science
Research Fund.


\begin{thebibliography}{99}
\bibitem{A2} Insofar as our goal is to extrapolate the finite-difference
  results to reproduce the linearly-induced magnetization, the $A^2$ term
  has no influence and may be neglected.
\end{thebibliography}
\end{document}